\documentclass[11pt]{article}
\usepackage{axodraw}
\usepackage{epsfig}
\input pix.sty

\newcommand{\us}[1]{\frac{1}{d-x}}

\newcommand{\tinymsbar}{{\overline{\mbox{\tiny\rm{MS}}}}}
\def\Ada(#1,#2)(#3,#4,#5){\DashCArc(#1,#2)(#3,#4,#5){3}}
\def\Lda(#1,#2)(#3,#4){\DashLine(#1,#2)(#3,#4){3}}

\def\ToprVBblob(#1,#2,#3,#4){\picb{#1(30,15)(15,-120,120)%
 #2(30,15)(15,120,240) #3(15,15)(15,60,300) #4(15,15)(15,-60,60)%
 \GCirc(45,15){2}{0}}}

\newcommand{\Lambdamsbar}{{\Lambda_\tinymsbar}}
\newcommand{\Nf}{N_{\rm f}}
\newcommand{\Nc}{N_{\rm c}}
\newcommand{\Tc}{T_{\rm c}}

\newcommand{\bmu}{\bar\mu}

\def\lsi{\raise0.3ex\hbox{$<$\kern-0.75em\raise-1.1ex\hbox{$\sim$}}}
\def\gsi{\raise0.3ex\hbox{$>$\kern-0.75em\raise-1.1ex\hbox{$\sim$}}}

\newcommand{\Tint}[1]{{\hbox{$\sum$}\!\!\!\!\!\!\int}_{\!\!\!\!#1}}
\newcommand{\rmii}[1]{{\mbox{\tiny\rm{#1}}}}
\def\Lwidth{1}

\def\Aegl(#1,#2)(#3,#4,#5){\PhotonArc(#1,#2)(#3,#4,#5){\Lwidth}
{6.283 #3 mul 360 div #4 #5 sub #4 #5 sub mul sqrt mul Ldensity mul}}
\def\Legl(#1,#2)(#3,#4){\Photon(#1,#2)(#3,#4){\Lwidth}
{#1 #3 sub #1 #3 sub mul #2 #4 sub #2 #4 sub mul add sqrt Ldensity mul}}
\def\ToprSBB(#1,#2,#3,#4,#5){\picb{#1(0,15)(7.5,15)  #1(37.5,15)(45,15)%
 #2(22.5,15)(15,0,70) #2(22.5,15)(15,110,180) #3(22.5,15)(15,180,360)%
 #4(22.5,30)(5,-10,190) #5(22.5,30)(5,190,350)}}
\def\ToprSBT(#1,#2,#3,#4){\picb{#1(0,15)(7.5,15)  #1(37.5,15)(45,15)%
 #2(22.5,15)(15,0,90) #2(22.5,15)(15,90,180) #3(22.5,15)(15,180,360)%
 #4(22.5,35)(5,-90,270)}}
\def\ToprSTB(#1,#2,#3,#4){\picb{#1(0,0)(22.5,0) #1(22.5,0)(45,0)%
 #2(22.5,15)(15,-90,70) #2(22.5,15)(15,110,270)%
 #3(22.5,30)(5,-10,190) #4(22.5,30)(5,190,350)}}
\def\ToprSTT(#1,#2,#3){\picb{#1(0,0)(22.5,0) #1(22.5,0)(45,0)%
 #2(22.5,15)(15,-90,90) #2(22.5,15)(15,90,270)%
 #3(22.5,35)(5,-90,270)}}

\makeatletter \@addtoreset{equation}{section} \makeatother
\renewcommand{\theequation}{\arabic{section}.\arabic{equation}}
\makeatletter
\renewcommand\section{\@startsection {section}{1}{\z@}%
                                   {-5.5ex \@plus -1ex \@minus -.2ex}
                                   {2.3ex \@plus.2ex}%
                                   {\normalfont\large\bfseries}}
\renewcommand\subsection{\@startsection{subsection}{2}{\z@}%
                                     {-3.25ex\@plus -1ex \@minus -.2ex}%
                                     {1.5ex \@plus .2ex}%
                                     {\normalfont\normalsize\bfseries}}
\renewcommand\thesection {\@arabic\c@section}
\renewcommand\thesubsection   {\thesection.\@arabic\c@subsection}
\renewcommand{\@seccntformat}[1]{%
\csname the#1\endcsname.\hspace{1.0em}}
\makeatother

\begin{document}

\begin{titlepage}
\begin{flushright}
BI-TP 2005/07\\
hep-ph/0503061\\
\end{flushright}
\begin{centering}
\vfill

{\Large{\bf Two-loop QCD gauge coupling at high temperatures}}

\vspace{0.8cm}

M.~Laine, 
Y. Schr\"oder 

\vspace{0.8cm}

{\em
Faculty of Physics, University of Bielefeld, 
D-33501 Bielefeld, Germany\\}

\vspace*{0.8cm}
 
\mbox{\bf Abstract}

\end{centering}

\vspace*{0.3cm}
 
\noindent
We determine the 2-loop effective gauge coupling of QCD at high
temperatures, defined as a matching coefficient appearing in the 
dimensionally reduced effective field theory. The result allows to 
improve on one of the classic non-perturbative probes for the convergence 
of the weak-coupling expansion at high temperatures, the comparison of full
and effective theory determinations of an observable called the spatial
string tension. We find surprisingly good agreement almost down to the 
critical temperature of the deconfinement phase transition. We also 
determine one new contribution of order $\mathcal{O}(g^6T^4)$ 
to the pressure of hot QCD.
\vfill
\noindent
 

\vspace*{1cm}
 
\noindent
March 2005

\vfill

\end{titlepage}

%
\section{Introduction}

Indirect signs for rapid thermalisation after heavy ion collisions 
at RHIC energies, derived for instance from the fact that hydrodynamic 
models assuming local thermodynamic equilibrium
appear to work very well~\cite{uh}, have  
underlined the need to understand the physics of thermal QCD 
at temperatures above a few hundred MeV. 

Given asymptotic freedom, a natural tool for these studies is the 
weak-coupling expansion~\cite{es}. Alas, it has been known since a long time 
that the weak-coupling expansion converges very slowly at all realistic 
temperatures~\cite{az,zk}. It also has theoretically a non-trivial structure, 
with odd powers of the gauge coupling~\cite{jk} and even coefficients that
can only be determined non-perturbatively~\cite{linde,gpy}.

On the other hand, the degrees of freedom responsible for the slow convergence
can be identified~\cite{bn,adjoint,gsixg}: 
they are the ``soft'' static 
colour-electric modes, parametrically $p \sim gT$ (leading to the 
odd powers in the gauge coupling), as well as the ``ultrasoft'' static 
colour-magnetic modes, parametrically $p \sim g^2T$ 
(leading to the non-perturbative coefficients 
in the weak-coupling expansion). Here $p$ denotes the characteristic 
momentum scale, $g$ the gauge coupling
and $T$ the temperature. The belief has been 
that perturbation theory restricted to 
parametrically hard scales $p \sim 2\pi T$ alone should
converge well, while the soft and the ultrasoft scales need to be 
treated either with ``improved'' analytic schemes, or then 
non-perturbatively. As a starting point for these demanding
tasks one may take, however, either the dimensionally reduced 
effective field theory~\cite{dr,generic} or
the hard thermal loop effective theory~\cite{bp}, which have been 
obtained by integrating out the parametrically hard scales.
 
Quantitative evidence for this picture can be obtained by choosing 
simple observables which can be determined reliably both 
with four-dimensional (4d) lattice 
simulations and with the soft/ultrasoft effective theory. This
forces us to restrict to static observables 
and, for the moment, mostly pure gauge theory. Various comparisons of this 
kind are summarised in Refs.~\cite{own,owe,chris}.
The most precise results are related to static correlation lengths 
in various quantum number channels~\cite{mu}, where good agreement 
has generally been found down to $T \sim 2 \Tc$, 
where $\Tc$ is the critical temperature of the deconfinement phase
transition. The thermodynamic pressure of QCD is also 
consistent with this picture~\cite{gsixg}, even though that comparison
is not unambiguous yet, due to the fact that the effective theory 
approach does not directly produce the physical number, but 
requires not-yet-determined ultraviolet matching 
coefficients for its interpretation~\cite{plaq}.\footnote{%
  For the status regarding a few other observables, see Refs.~\cite{mv,ag,bmp}.
  } 

The purpose of this paper is to study another observable for which
an unambiguous comparison is possible. The observable is the 
``spatial string tension'', $\sigma_s$. 4d lattice 
determinations of $\sigma_s$ in pure SU(3) gauge theory exist 
since a while already~\cite{boyd} but, as has most recently
been stressed in Ref.~\cite{pg2}, the comparison 
with effective theory results shows a clear discrepancy.  
In order to improve on the 
resolution on the effective theory side, we compute here
the gauge coupling of the dimensionally reduced 
theory up to 2-loop order. Combining with other ingredients~\cite{mt,pg},
to be specified below, allows then for a precise comparison. We find
that once the 2-loop corrections are included, the match to 4d lattice data 
improves quite significantly and supports the picture outlined above. 

The plan of this paper is the following.
In \se\ref{se:gE2} we present the 2-loop computation of the  
effective gauge coupling of the dimensionally reduced theory. 
In \se\ref{se:num} we discuss the numerical evaluation of this result. 
In \se\ref{se:string} we use the outcome for estimating the spatial 
string tension, and compare with 4d lattice data.
We conclude in \se\ref{se:concl}.

%
\section{Effective gauge coupling}
\la{se:gE2}

We consider finite temperature QCD with the gauge group 
SU($\Nc$), and $\Nf$ flavours of massless quarks. In dimensional 
regularisation the bare Euclidean
Lagrangian reads, before gauge fixing,   
\ba
 S_\rmi{QCD} & = & \int_0^{\beta} \! {\rm d}\tau \int \! {\rm d}^d x\, 
 {\cal L}_\rmi{QCD}, \\
 {\cal L}_\rmi{QCD} & = & 
 \fr14 F_{\mu\nu}^a F_{\mu\nu}^a+
 \bar\psi \gamma_\mu D_\mu \psi 
 \;, \la{SQCD}
\ea
where $\beta = T^{-1}$,  $d=3-2\epsilon$, $\mu,\nu=0,...,d$, 
$F_{\mu\nu}^a = \partial_\mu A_\nu^a - \partial_\nu A_\mu^a + 
g_B f^{abc} A_\mu^b A_\nu^c$, 
$D_\mu = \partial_\mu - i g_B A_\mu$, 
$A_\mu = A_\mu^a T^a$, 
$T^a$
are Hermitean generators of SU($\Nc$) normalised such that 
$\tr [ T^a T^b ] = \delta^{ab}/2$, 
$\gamma_\mu^\dagger = \gamma_\mu$, 
$\{\gamma_\mu,\gamma_\nu\} = 2 \delta_{\mu\nu}$,
$g_B$ is the bare gauge coupling,  
and $\psi$ carries Dirac, colour, and flavour indices.   
We use the standard symbols 
$C_A = \Nc, C_F = (\Nc^2 - 1)/(2 \Nc), T_F = \Nf/2$
for the various group theory factors emerging.

At high enough temperatures, the dynamics of \eq\nr{SQCD} 
is contained in a simpler, dimensionally reduced effective 
field theory~\cite{dr,generic,bn}: 
\ba
 S_\rmi{EQCD} & = &  \int \! {\rm d}^d x\, {\cal L}_\rmi{EQCD}, \\
 {\cal L}_\rmi{EQCD} & = & 
 \fr14 F_{ij}^a F_{ij}^a +
 \tr [D_i,B_0]^2 + 
 m_\rmi{E}^2\tr [ B_0^2 ] +\lambda_\rmi{E}^{(1)} (\tr [B_0^2])^2
 +\lambda_\rmi{E}^{(2)} \tr [B_0^4] + ...\; . 
 \hspace*{0.5cm} \la{EQCD}
\ea
Here $i=1,...,d$, $F_{ij} = \partial_i B_j^a - \partial_j B_i^a + 
g_\rmi{E} f^{abc} B_i^b B_j^c$, and
$D_i = \partial_i - i g_\rmi{E} B_i$. 
The fields $B^a_\mu$ have the dimension $[\mbox{GeV}]^{1/2-\epsilon}$, 
due to a trivial rescaling with $T^{1/2}$. 
Note also that the 
quartic couplings $\lambda_\rmi{E}^{(1)}$, $\lambda_\rmi{E}^{(2)}$
are linearly dependent for $N_c \le 3$, since then 
$\tr [B_0^4] = \fr12 (\tr[B_0^2])^2$.

The theory in \eq\nr{EQCD}
has been truncated to be super-renormalisable; 
that is, higher order operators~\cite{sc} (see also Refs.~\cite{mrs,do}
and references therein) 
have been dropped. The relative error thus induced has been 
discussed for generic Green's functions in Ref.~\cite{parity}, and 
for the particular case of the pressure of hot QCD in Ref.~\cite{gsixg}. 
In the following we concentrate on an observable 
dynamically determined by the 
colour-magnetic scale $p\sim g^2 T$, 
and it is easy to see that in this case the higher
order operators do not play any role at the order we are working. 

The effective parameters in \eq\nr{EQCD} can be determined by 
matching, that is, by requiring that QCD and EQCD 
produce the same results, within the domain of validity of the latter theory. 
It is essential that infrared (IR) physics is treated in the same
way in both theories at the matching stage and, as outlined in 
Ref.~\cite{bn}, the most convenient implementation of this requirement is 
to perform computations on both sides using ``unresummed'' propagators. 
We follow this procedure here. 

The matching simplifies further by using the background field 
gauge (Ref.~\cite{lfa} and references therein). As this is 
essential for what follows, we start by briefly recalling 
the basic advantage of this approach. For a concise yet
rigorous overview of the technique, see Ref.~\cite{lw}.

We denote the background gauge potential with $B^a_\mu$, 
and the gauge-invariant combination following from 
$F_{\mu\nu}^a(B) F_{\mu\nu}^a(B)$ symbolically as $B^2 + g B^3 + g^2 B^4$.
Now, the computation of the effective Lagrangian by integrating
out the hard scales $p\sim 2\pi T$ produces, in general, 
an expression of the type
\be
 \mathcal{L}_\rmi{eff} \sim c_2\, B^2 + c_3\, g B^3 + c_4\, g^2 B^4 + ...
 \;, \la{Leff1}
\ee
where $c_i$ are coefficients of the form $c_i = 1 + \mathcal{O}(g^2)$.
As the next step we are free to define 
a canonically normalised effective field $B_\rmi{eff}$ 
as $B_\rmi{eff}^2 \equiv c_2 B^2$. Then the effective Lagrangian obtains
the form
\be
 \mathcal{L}_\rmi{eff} \sim 
 B_\rmi{eff}^2 +  {c_3} { c_2^{-3/2} } \, g B_\rmi{eff}^3 + 
 c_4 c_2^{-2} \, g^2 B_\rmi{eff}^4 + ...
 \;. \la{Beff}
\ee
We can now read off the effective gauge coupling from the gauge-invariant
structure: 
\be
 g_\rmi{eff} = c_3 c_2^{-3/2} \, g  =  c_4^{1/2} c_2^{-1}  \, g
 \;. \la{geff0}
\ee
We observe that two independent computations are needed 
for the determination of $g_\rmi{eff}$, but we can 
choose whether to go through the 3-point or the 4-point 
function, in addition to the 2-point function 
(that is, using $c_3$ or $c_4$, in addition to $c_2$). 

The background field gauge economises this setup. Indeed, 
the effective action is then gauge-invariant not only in terms of 
$B_\rmi{eff}$, but also in terms of the original field $B$~\cite{lfa}.
Writing \eq\nr{Leff1} as
\be
 \mathcal{L}_\rmi{eff} \sim c_2
 \Bigl[
  B^2 + c_3 c_2^{-1}  \, g B^3 + c_4 c_2^{-1}  \, g^2 B^4  
 \Bigr] + ...
 \;, 
\ee
gauge invariance in terms of $B$ now tells us that 
$
 c_3 = c_2
$ and 
$
  c_4 = c_2.
$
Combining with \eq\nr{geff0}, we obtain
\be
 g_\rmi{eff} = c_2^{-1/2} \; g
 \;, \la{geff}
\ee
so that it is enough to carry out one single 2-point computation, 
in order to obtain $g_\rmi{eff}$. In our case, the role of 
$g_\rmi{eff}$ is played by $g_\rmi{E}$ (cf.\ \eq\nr{EQCD}).

The class of background field gauges still allows for a general (bare) 
gauge parameter, $\xi$. As a cross-check we have carried out all 
computations with a general $\xi$, and verified that it cancels at the end. 
To be definite, we denote $(\xi)_\rmi{here} = 1 - (\xi)_\rmi{standard}$, 
so that the gauge field propagator reads
\be
 \Bigl\langle A^a_\mu(q) A^b_\nu(-q)
 \Bigr\rangle 
 = \delta^{ab}
 \biggl[
 \frac{\delta_{\mu\nu}}{q^2} - \xi \frac{q_\mu q_\nu}{(q^2)^2} 
 \biggr]
 \;.
\ee  

\begin{figure}[t]
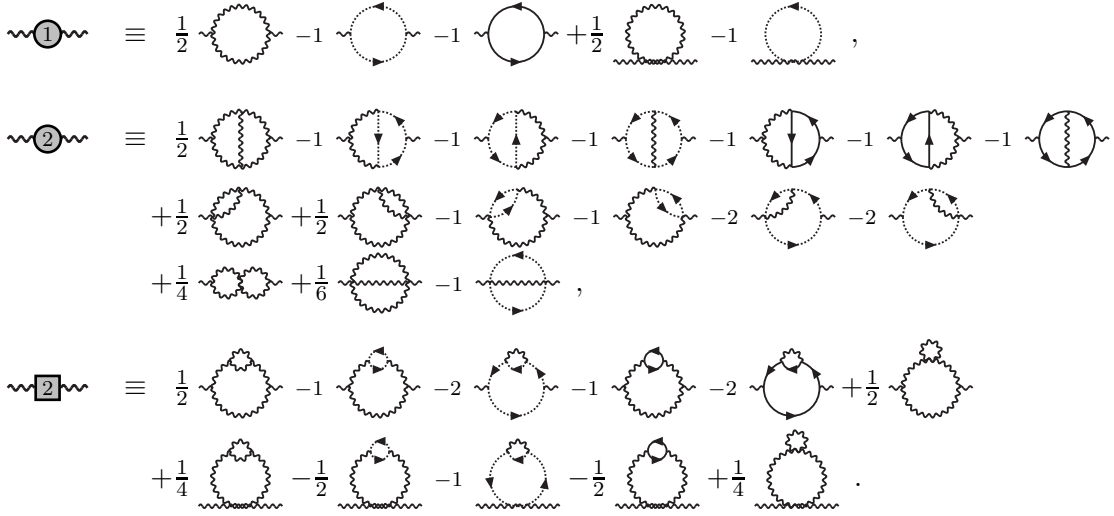


\begin{eqnarray*}
%
%
\TopoS(\Lgl) & \equiv &
\sy{}12 \TopoSB(\Lgl,\Agl,\Agl)
\sm{-1} \TopoSB(\Lgl,\Agh,\Agh)
\sm{-1} \TopoSB(\Lgl,\Aqu,\Aqu)
\sy+12 \TopoST(\Lgl,\Agl)
\sm{-1} \TopoST(\Lgl,\Agh) \;,
\\[0ex]
&& \nn[0ex]
%
%
\ToptSi(\Lgl) & \equiv & 
\sy{}12 \ToptSM(\Legl,\Agl,\Agl,\Agl,\Agl,\Lgl)
\sm{-1} \ToptSM(\Legl,\Agh,\Agl,\Agl,\Agh,\Lgh)
\sm{-1} \ToptSM(\Legl,\Agl,\Agh,\Agh,\Agl,\Lagh)
\sm{-1} \ToptSM(\Legl,\Agh,\Agh,\Agh,\Agh,\Lgl)
\sm{-1} \ToptSM(\Legl,\Aqu,\Agl,\Agl,\Aqu,\Lqu)
\sm{-1} \ToptSM(\Legl,\Agl,\Aqu,\Aqu,\Agl,\Laqu)
\sm{-1} \ToptSM(\Legl,\Aqu,\Aqu,\Aqu,\Aqu,\Lgl) \nn[0ex]&&{} \hspace*{-0.3cm}
\sy+12 \ToptSAl(\Legl,\Agl,\Agl,\Agl,\Agl)
\sy+12 \ToptSAr(\Legl,\Agl,\Agl,\Agl,\Agl) 
\sm{-1} \ToptSAl(\Legl,\Agl,\Agh,\Agl,\Agh)
\sm{-1} \ToptSAr(\Legl,\Agh,\Agl,\Agl,\Agh) 
\sm{-2} \ToptSAl(\Legl,\Agh,\Agh,\Agh,\Agl)
\sm{-2} \ToptSAr(\Legl,\Agh,\Agh,\Agh,\Agl)  \nn[0ex]&&{} \hspace*{-0.3cm}
\sy+14 \ToptSE(\Lgl,\Agl,\Agl,\Agl,\Agl)
\sy+16 \ToptSS(\Lgl,\Agl,\Agl,\Lgl) 
\sm{-1} \ToptSS(\Lgl,\Agh,\Agh,\Lgl) \;,
\\[0ex]
&& \nn[0ex]
%
%
\ToptSr(\Lgl ) & \equiv &
\sy{}12 \ToprSBB(\Legl,\Agl,\Agl,\Agl,\Agl)
\sm{-1} \ToprSBB(\Legl,\Agl,\Agl,\Agh,\Agh)
\sm{-2} \ToprSBB(\Legl,\Agh,\Agh,\Agl,\Aagh)
\sm{-1} \ToprSBB(\Legl,\Agl,\Agl,\Aqu,\Aqu)
\sm{-2} \ToprSBB(\Legl,\Aqu,\Aqu,\Agl,\Aaqu)
\sy+12 \ToprSBT(\Legl,\Agl,\Agl,\Agl) \nn[1ex]&&{} \hspace*{-0.3cm}
\sy+14 \ToprSTB(\Legl,\Agl,\Agl,\Agl)
\sy-12 \ToprSTB(\Legl,\Agl,\Agh,\Agh) 
\sm{-1}\ToprSTB(\Legl,\Agh,\Agl,\Aagh)
\sy-12 \ToprSTB(\Legl,\Agl,\Aqu,\Aqu) 
\sy+14 \ToprSTT(\Legl,\Agl,\Agl)  \;.
\end{eqnarray*}

\caption[a]{\it The 1-loop and 2-loop self-energy diagrams
in the background field gauge. Wavy lines represent  
gauge fields, dotted lines ghosts,  and solid lines fermions. 
The 2-loop graphs have been divided into two-particle-irreducible 
and two-particle-reducible contributions.}  
\label{fig:2pt}

\end{figure}

In order to match the effective gauge coupling, we need to compute
the 2-loop gluon self-energy, $\Pi_{\mu\nu}(p)$, for the background gauge
potential $B^a_\mu$. The graphs entering are shown in \fig\ref{fig:2pt}.
The external momentum $p$ is taken purely spatial, 
$p = (0,\vec{p})$, while the heat bath is timelike, with 
Euclidean four-velocity $u = (1,0)$, so that $u\cdot u = 1, u \cdot p = 0$. 
In this case $\Pi_{\mu\nu}$ 
has three independent components ($\Pi_{0i}$, $\Pi_{i0}$ vanish identically), 
\be
 \Pi_{00}(\vec{p}) \equiv \Pi_\rmi{E}(\vec{p}^2) 
 \;, \quad
 \Pi_{ij}(\vec{p}) \equiv 
 \biggl(
 \delta_{ij} - \frac{p_i p_j}{\vec{p}^2}
 \biggr) \Pi_\rmi{T}(\vec{p}^2) 
 + \frac{p_i p_j}{\vec{p}^2} \Pi_\rmi{L}(\vec{p}^2)
 \;, \la{Pidef}
\ee
where $i,j = 1,...,d$. In fact loop corrections to 
the spatially longitudinal part $\Pi_\rmi{L}$ also vanish, 
so that only two non-trivial functions, $\Pi_\rmi{E}, \Pi_\rmi{T}$,  remain.

Since we are carrying out a matching computation, any possible IR divergences
cancel as we subtract the contribution of EQCD. Therefore we may Taylor-expand
$\Pi_{\mu\nu}(p)$ to second order in $\vec{p}^2$. This leads to the nice 
simplification that the results on the EQCD side vanish 
identically in dimensional
regularization, due to the absence of any mass scales in the propagators.
Thus we only need to compute unresummed integrals on the QCD side.

After the Taylor-expansion, 
the 2-loop QCD integrals can all be cast in the form
\be
 I(i_1,i_2;j_1,j_2,j_3;k_1,k_2,k_3) \equiv 
 \Tint{q,r} \frac{q_0^{i_1} r_0^{i_2} 
 (\vec{q}\cdot\vec{p})^{j_1}
 (\vec{r}\cdot\vec{p})^{j_2}
 (\vec{q}\cdot\vec{r})^{j_3}
 }{[q_0^2 + \vec{q}^2]^{k_1}[r_0^2 + \vec{r}^2]^{k_2}
 [(q_0+r_0)^2 + (\vec{q}+\vec{r})^2]^{k_3}}
 \;. \la{int}
\ee 
The indices here are non-negative integers, and the measure
is the standard Matsubara sum-integral (bosonic or fermionic), 
with the spatial part 
$\int\! {\rm d}^d \vec{q}/(2\pi)^d \int\! {\rm d}^d \vec{r}/(2\pi)^d$. 

To reduce integrals of the type in~\eq\nr{int} to a small 
set of ``master integrals'', we employ symmetries following from
exchanges of integration variables, as well as general partial
integration identities for the spatial parts of the momentum 
integrations. The implementation of these identities follows
the procedure outlined by Laporta~\cite{laporta}, 
in analogy with Ref.~\cite{ysproc}. We are lead both to very 
simple 1-loop recursion relations, such as 
\be
 I(2 i_1,0;0,0,0;k_1,1,0) =
 \frac{2k_1 - 2 - d}{2k_1 - 2} 
 I(2 i_1-2,0;0,0,0;k_1-1,1,0)
 \;, 
\ee 
as well as well-known but less obvious 2-loop ones~\cite{ae}, like 
\be
 I(0,0;0,0,0;1_\rmi{b},1_\rmi{b},1_\rmi{b}) = 0 
 \;,
\ee
where the subscripts refer to bosonic four-momenta. 

After this reduction, only six master integrals remain:
\be
 I_\rmi{b}(n) = \Tint{q_\rmi{b}} \frac{1}{(q^2)^n} 
 \;, \quad
 I_\rmi{f}(n) = \Tint{q_\rmi{f}} \frac{1}{(q^2)^n} 
 \;,
\ee
where $q_\rmi{b},q_\rmi{f}$ refer to bosonic and fermionic Matsubara
momenta, respectively, and $n=1,2,3$. For a vanishing quark chemical 
potential, as we assume to be the case here, 
the fermionic integrals reduce further to the bosonic ones, 
\be
 I_\rmi{f}(n) = 
 \Bigl( 
  2^{2n - d} - 1 
 \Bigr) I_\rmi{b}(n)
 \;,
\ee
leaving only three master integrals. They are known explicitly, 
\be 
 I_\rmi{b}(n) = \frac{2 \pi^{d/2} T^{1+d}}{(2 \pi T)^{2 n}}
 \frac{\Gamma(n-d/2)}{\Gamma(n)} \zeta(2n - d)
 \;. \la{Imaster}
\ee
This expression is easily expanded in $\epsilon$ and, in the following, 
we need terms up to $\mathcal{O}(\epsilon)$. For completeness, the 
relevant expansions are shown in Appendix A. 

Writing now the
Taylor-expanded bare 2-point function $\Pi_\rmi{T}$
of \eq\nr{Pidef} as
\ba
 \Pi_\rmi{T}(\vec{p}^2) & \equiv &  
 \Pi_\rmi{T}(0) + \vec{p}^2  
 \Pi_\rmi{T}'(0) + ... \nn 
 & \equiv & 
 \sum_{n=1}^\infty \Pi_\rmi{Tn}(0)(g_B^2)^n +  
 \vec{p}^2
 \sum_{n=1}^\infty \Pi_\rmi{Tn}'(0)(g_B^2)^n +  
 ...
 \;, 
\ea
where $g_B$ is the bare gauge coupling, 
and correspondingly for $\Pi_\rmi{E}$, our results read
\ba
 \Pi_\rmi{T1}(0) & = & 0 
  \;,  \\
 \Pi_\rmi{T1}'(0) & = & 
 \frac{d-25}{6} C_A I_\rmi{b}(2) + \fr43 T_F I_\rmi{f}(2) 
 \;, \la{PiT1p} \\
 \Pi_\rmi{T2}(0) & = & 0
 \;, \\
 \Pi_\rmi{T2}'(0) & = & 
 \frac{(d-3)(d-4)}{d(d-2)(d-5)(d-7)}
 \biggl\{
  2 (4d^2 - 21 d - 7) C_A^2 I_\rmi{b}^2(2) -
 \nn & & 
 \hphantom{  \frac{(d-3)(d-4)}{d(d-2)(d-5)(d-7)}
 \biggl\{ } 
  -  
  8 \Bigl[ 4 C_F + (d^2 - 6d + 1) C_A
    \Bigr] T_F I_\rmi{b}(2) I_\rmi{f}(2) - 
 \nn & & 
 \hphantom{  \frac{(d-3)(d-4)}{d(d-2)(d-5)(d-7)}
 \biggl\{ } 
 - 
 \Bigl[
  (d^3 - 12 d^2 + 39 d - 12) C_A -
 \nn & & 
 \hphantom{  \frac{(d-3)(d-4)}{d(d-2)(d-5)(d-7)}
 \biggl\{ - \Bigl[ } 
 - 2 (d^3 - 12 d^2 +41 d - 14) C_F 
 \Bigr] T_F I_\rmi{f}^2(2)  
 \biggr\} + 
 \nn & & + 
 \frac{(d-1)}{3 d(d-7)}
 \biggl\{
   (d^2-31 d + 144) 
    \Bigl[ 
      4 T_F I_\rmi{f}(1) - (d-1) C_A I_\rmi{b}(1)
    \Bigr] C_A I_\rmi{b}(3) - 
  \nn & & 
   \hphantom{ \frac{(d-1)}{3 d(d-7)}
   \biggl\{ }
   - 8 (d-1)(d-6) C_F T_F
    \Bigl[ I_\rmi{b}(1) - I_\rmi{f}(1) 
    \Bigr] I_\rmi{f}(3)
 \biggr\}
 \;, \la{PiT2p} \\
 \Pi_\rmi{E1}(0) & = & 
 -(d-1) \Bigl[ 
  4 T_F I_\rmi{f}(1) - 
  (d-1) C_A I_\rmi{b}(1) \Bigr] 
 \;, \la{PiE1} \\ 
 \Pi_\rmi{E1}'(0) & = & 
 - \biggl[ \frac{d^2 -  5 d +28 }{6} +  (d -3) \xi \biggr] C_A I_\rmi{b}(2)
 + \frac{2(d-1)}{3} T_F I_\rmi{f}(2) 
 \;, \la{PiE1p} \\
 \Pi_\rmi{E2}(0) & = & 
 (d-1)(d-3)
 \biggl\{
   (1+\xi) \Bigl[
    4 T_F I_\rmi{f}(1) - (d-1) C_A I_\rmi{b}(1) 
   \Bigr] C_A I_\rmi{b}(2) +  
 \nn & & \hphantom{ (d-1)(d-3)
 \biggl\{ }
 + 4 C_F T_F \Bigl[
   I_\rmi{b}(1) - I_\rmi{f}(1) 
  \Bigr] I_\rmi{f}(2)
 \biggr\}
 \;. \la{PiE2}
\ea
We leave out the lengthy expression for $\Pi_\rmi{E2}'(0)$, 
as it is not needed in the following. 

The bare results
need still  to be renormalised. The bare gauge coupling is written
as $g_B^2 = g^2(\bmu) Z_g$, where $g^2(\bmu)$ is the renormalised 
gauge coupling, 
$\bmu$ is an $\msbar$ scheme scale parameter introduced 
through $\mu^2 \equiv \bmu^2 e^{\gamma_\rmi{E}}/4\pi$,
and the combination $\mu^{-2\epsilon} g^2(\bmu)$ 
is dimensionless. Denoting
\ba
 \beta_0 & \equiv &
 \frac{-22 C_A + 8 T_F}{3}
 \;, \\
 \beta_1 & \equiv & 
 \frac{-68 C_A^2 + 40 C_A T_F + 24 C_F T_F}{3}
 \;, 
\ea
the factor $Z_g$ reads
\be
 Z_g = 1 + 
 \frac{1}{(4\pi)^2} \frac{\beta_0}{2\epsilon}
 \mu^{-2\epsilon}g^2(\bmu) + 
 \frac{1}{(4\pi)^4}
 \biggl[
  \frac{\beta_1}{4\epsilon} + \frac{\beta_0^2}{4 \epsilon^2}
 \biggr] 
 \mu^{-4\epsilon}g^4(\bmu) + 
 \mathcal{O}(g^6)
 \;, \la{Zg}
\ee
and the renormalised gauge coupling satisfies, 
in the limit $\epsilon\to 0$,
\be
 \bmu \frac{{\rm d}}{{\rm d}\bmu} g^2(\bmu) = 
 \frac{\beta_0}{(4\pi)^2} g^4(\bmu) + 
 \frac{\beta_1}{(4\pi)^4} g^6(\bmu) + 
 \mathcal{O}(g^8)
 \;. \la{rge}
\ee

To proceed, we first cross-check
our results for $\Pi_\rmi{E}$ against known expressions.  
After the fields $B_0^a$ of EQCD are normalised to their 
canonical form (cf.\ \eq\nr{Beff}), 
$(B_0^a B_0^a)_\rmi{E} 
  \equiv 
(B_0^a B_0^a)_\rmi{4d} 
 [1 + \Pi_\rmi{E1}'(0)] / T$,  we 
obtain for the matching coefficient $m_\rmi{E}^2$, 
\be
 m_\rmi{E}^2 = g_B^2 \, \Pi_\rmi{E1}(0) + 
 g_B^4 \, \Bigl[
  \Pi_\rmi{E2}(0) - \Pi_\rmi{E1}'(0) \Pi_\rmi{E1}(0)
 \Bigr]
 + \mathcal{O}(g_B^6)
 \;. 
\ee 
Inserting \eqs\nr{PiE1}--\nr{PiE2}, 
the $\xi$-dependence duly cancels. Re-expanding also $g_B^2$ in 
terms of the renormalised gauge coupling, 
and writing then~\cite{gsixg}
\be
 m_\rmi{E}^2 \equiv T^2
 \biggl\{
  g^2(\bmu) \Bigl[ 
   \alpha_\rmi{E4} + \alpha_\rmi{E5} \epsilon 
 \Bigr] + 
 \frac{g^4(\bmu)}{(4\pi)^2}
 \Bigl[ 
   \alpha_\rmi{E6} + \beta_\rmi{E2} \epsilon
 \Bigr] + \mathcal{O}(g^6,\epsilon^2)
 \biggr\}
 \;, 
\ee
we recover the known values of $\alpha_\rmi{E4}$, 
$\alpha_\rmi{E5}$ and $\alpha_\rmi{E6}$~\cite{gsixg}
(for original derivations, see Ref.~\cite{bn}
and references therein). We also obtain
\ba
 \beta_\rmi{E2} & = & 
 \frac{1}{36} C_A^2
 \biggl\{
 264 \ln^2\biggl( \frac{\bmu e^{\gamma_\rmi{E}}}{4\pi T} \biggr) + 
 \biggl[
  80 - 176 \gamma_\rmi{E} + 176 \frac{\zeta'(-1)}{\zeta(-1)} 
 \biggr] \ln \biggl( \frac{\bmu e^{\gamma_\rmi{E}}}{4\pi T} \biggr) + 
 \nn & & \hspace*{1cm} 
 + 8 + 11\pi^2 - 88 \gamma_\rmi{E}^2 - 40 \gamma_\rmi{E} - 176 \gamma_1 + 
 40 \frac{\zeta'(-1)}{\zeta(-1)}
 \biggr\} + 
 \nn & + & 
 C_F T_F 
 \biggl\{ 
 -8 \ln \biggl( \frac{\bmu e^{\gamma_\rmi{E}}}{4\pi T} \biggr)
 - 2 - \frac{20}{3} \ln 2+ 4 \gamma_\rmi{E} - 4 \frac{\zeta'(-1)}{\zeta(-1)}
 \biggr\} +  
 \nn & + &
 \frac{1}{36} C_A T_F 
 \biggl\{ 
 168 \ln^2 \biggl( \frac{\bmu e^{\gamma_\rmi{E}}}{4\pi T} \biggr)
 + \biggl[
    232 - 432 \ln 2-112 \gamma_\rmi{E} + 112 
   \frac{\zeta'(-1)}{\zeta(-1)}
 \biggr] 
  \ln \biggl( \frac{\bmu e^{\gamma_\rmi{E}}}{4\pi T} \biggr) + 
 \nn & & \hspace*{1cm}
 + 28 + 7 \pi^2 
 + 24 \ln 2 - 64 \ln^2 2
 - 56 \gamma_\rmi{E}^2 - 72 \gamma_\rmi{E} 
 + 128 \gamma_\rmi{E} \ln 2 
 - 112 \gamma_1 + 
 \nn & & \hspace*{1cm} 
 + 72
   \frac{\zeta'(-1)}{\zeta(-1)}
 -128 \ln 2
   \frac{\zeta'(-1)}{\zeta(-1)}
 \biggr\} + 
 \nn & + & \fr19 T_F^2 
 \biggl\{ 
  -24 \ln^2 
  \biggl( \frac{\bmu e^{\gamma_\rmi{E}}}{4\pi T} \biggr)
 + \biggl[
 8 - 48 \ln 2+ 16 \gamma_\rmi{E} - 16 
   \frac{\zeta'(-1)}{\zeta(-1)}
 \biggr]
  \ln \biggl( \frac{\bmu e^{\gamma_\rmi{E}}}{4\pi T} \biggr) + 
 \nn & & \hspace*{1cm} 
 + 4 - \pi^2 - 8 \ln 2+16\ln^2 2
 + 8 \gamma_\rmi{E}^2 -8 \gamma_\rmi{E}
 + 32 \gamma_\rmi{E} \ln 2 + 16 \gamma_1 + 
 \nn & & \hspace*{1cm} 
 + 8
   \frac{\zeta'(-1)}{\zeta(-1)}
 -32 \ln 2
   \frac{\zeta'(-1)}{\zeta(-1)}
 \biggr\} 
 \;. \la{bE2}
\ea
Here $\gamma_1$ is a Stieltjes constant, defined through
the series $\zeta(s) = 1/(s-1) + 
\sum_{n = 0}^\infty \gamma_n (-1)^n (s - 1)^n/n!$.
(Note that the Euler gamma-constant is $\gamma_\rmi{E} \equiv \gamma_0$.)
The result in \eq\nr{bE2}, first obtained in Ref.~\cite{ysproc2} 
by employing the results of Ref.~\cite{bn}, 
contributes to the pressure of hot QCD at $\mathcal{O}(g^6T^4)$~\cite{gsixg}.
We rewrite the expression here, since Ref.~\cite{ysproc2}  employed
an extremely compactified notation.

We then move to consider the transverse spatial part, $\Pi_\rmi{T}(\vec{p}^2)$.
According to~\eq\nr{geff}, 
this directly determines the effective gauge coupling:
\be 
 g_\rmi{E}^2 = T \Bigl\{ g_B^2 - g_B^4 \, \Pi_{T1}'(0) + 
 g_B^6 \, \Bigl[
 \Bigl( \Pi_{T1}'(0) \Bigr)^2 - \Pi_{T2}'(0) 
 \Bigr] + \mathcal{O}(g_B^8) \Bigr\}
 \;. 
\ee
Re-expanding again in terms of $g^2(\bmu)$, 
we parameterise the result (following Ref.~\cite{gsixg}) as
\ba
 g_\rmi{E}^2 & \equiv & 
 T \biggl\{
 g^2(\bmu) + 
 \frac{g^4(\bmu)}{(4\pi)^2}
 \Bigl[
 \alpha_\rmi{E7} + \beta_\rmi{E3} \epsilon + \mathcal{O}(\epsilon^2) 
 \Bigr] + 
 \frac{g^6(\bmu)}{(4\pi)^4}
 \Bigl[ 
  \gamma_\rmi{E1} + \mathcal{O}(\epsilon)
 \Bigr] + \mathcal{O}(g^8) \biggr\}
 \;. \hspace*{0.7cm} \la{gE1}
\ea
We recover the known expression~\cite{hl,generic} for $\alpha_\rmi{E7}$, 
\be
 \alpha_\rmi{E7} = 
 -\beta_0 \ln  \biggl( \frac{\bmu e^{\gamma_\rmi{E}}}{4\pi T} \biggr) 
 + \fr13 C_A - \frac{16}{3} T_F \ln 2
 \;,  \la{gE2}
\ee
and obtain the new contributions
\ba
 \beta_\rmi{E3} & = &
 \frac{1}{12} C_A 
 \biggl[
  88 \ln^2  \biggl( \frac{\bmu e^{\gamma_\rmi{E}}}{4\pi T} \biggr) 
 + 8 \ln \biggl( \frac{\bmu e^{\gamma_\rmi{E}}}{4\pi T} \biggr)
 + 11 \pi^2 - 88 \gamma_\rmi{E}^2 - 176 \gamma_1
 \biggr] - 
 \nn & - & 
 \fr13 T_F
 \biggl[ 
  8 \ln^2 \biggl( \frac{\bmu e^{\gamma_\rmi{E}}}{4\pi T} \biggr) + 
  32 \ln 2 \ln \biggl( \frac{\bmu e^{\gamma_\rmi{E}}}{4\pi T} \biggr)
 + \pi^2 + 16 \ln^2 2 - 8 \gamma_\rmi{E}^2 - 16 \gamma_1
 \biggr]
 \;,  \la{bE3} \\
 \gamma_\rmi{E1} & = & 
 -\beta_1 \ln  \biggl( \frac{\bmu e^{\gamma_\rmi{E}}}{4\pi T} \biggr) 
 + \biggl[
 \beta_0 \ln  \biggl( \frac{\bmu e^{\gamma_\rmi{E}}}{4\pi T} \biggr) 
 - \fr13 C_A + \frac{16}{3} T_F \ln 2
 \biggr]^2 - 
 \nn & -  & 
 \frac{1}{18} \biggl\{ 
 C_A^2 \Bigl[ -341 + 20 \zeta(3) \Bigr] + 
 4 C_A T_F \Bigl[
 43 + 24 \ln 2 + 5 \zeta(3)
 \Bigr] +
 \nn & & \hspace*{0.5cm}
 + 6 C_F T_F \Bigl[
 23 + 80 \ln 2 - 14 \zeta(3) 
 \Bigr] \biggr\} 
 \;. \la{gE3}
\ea
The first one, $\beta_\rmi{E3}$, constitutes again an 
$\mathcal{O}(g^6T^4)$ contribution to the pressure of hot QCD~\cite{gsixg}, 
while the latter one is the desired finite 2-loop correction to the 
effective gauge coupling.

%
\section{Numerical evaluation}
\la{se:num}

We wish to compare numerically the 1-loop and 2-loop expressions
for $g_\rmi{E}^2$, in the limit $\epsilon\to 0$. 
When carrying out such a comparison, it is important
to specify the definitions of the $\Lambdamsbar$-parameters. 
Following standard procedures, we solve \eq\nr{rge} exactly at 2-loop level, 
and define
\be
 \Lambdamsbar \equiv \lim_{\bmu\to\infty}
 \bmu \Bigl[ b_0 g^2(\bmu) \Bigr] ^{-b_1/2 b_0^2}
 \exp \Bigl[ -\frac{1}{2 b_0 g^2(\bmu)}\Bigr]
 \;, \la{Lamdef}
\ee
where $b_0 \equiv - \beta_0/2 (4\pi)^2$, 
$b_1 \equiv -\beta_1/2 (4\pi)^4$. 
For large $\bmu$ this leads to the usual behaviour
\be
 \frac{1}{g^2(\bmu)} \approx 
 2 b_0 \ln\frac{\bmu}{\Lambdamsbar}
 + \frac{b_1}{b_0} \ln\biggl(
 2 \ln\frac{\bmu}{\Lambdamsbar}
 \biggr)
 \;. \la{Lambdamsbar}
\ee
In the 1-loop case, we set $b_1 \equiv 0$
in \eqs\nr{Lamdef}, \nr{Lambdamsbar}.

\begin{figure}[t]


\centerline{%
\epsfysize=5.0cm\epsfbox{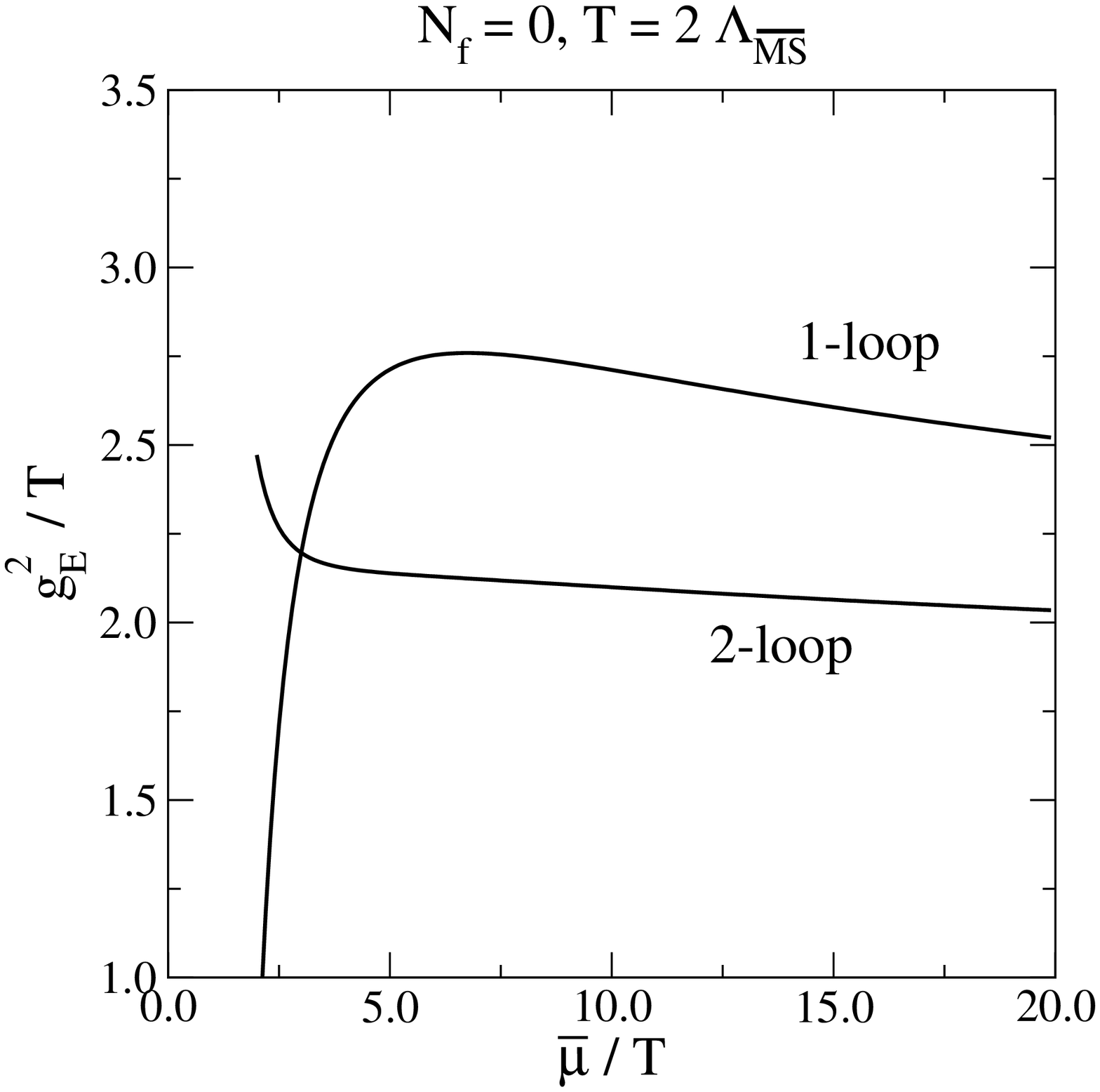}%
~~\epsfysize=5.0cm\epsfbox{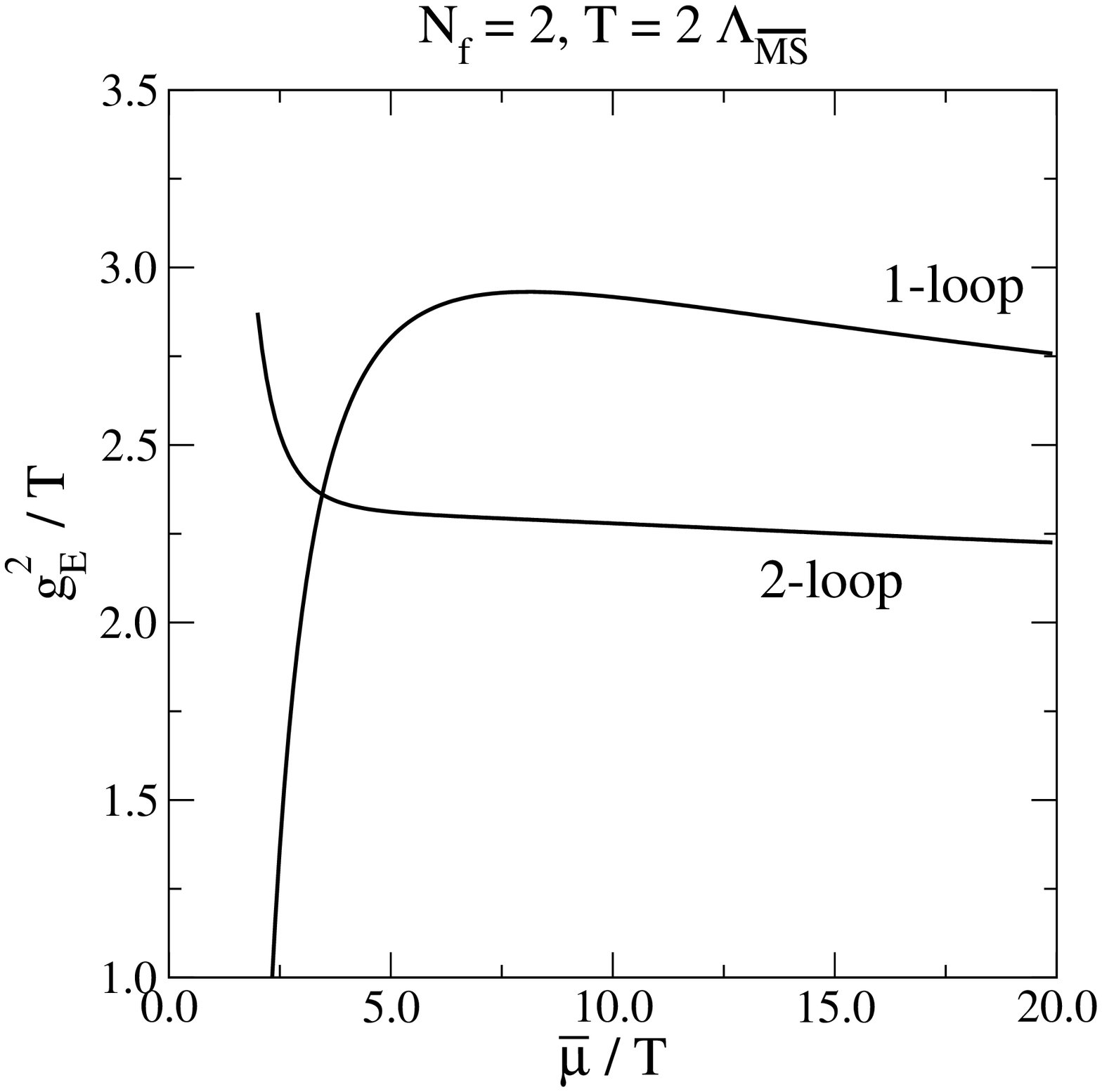}%
~~\epsfysize=5.0cm\epsfbox{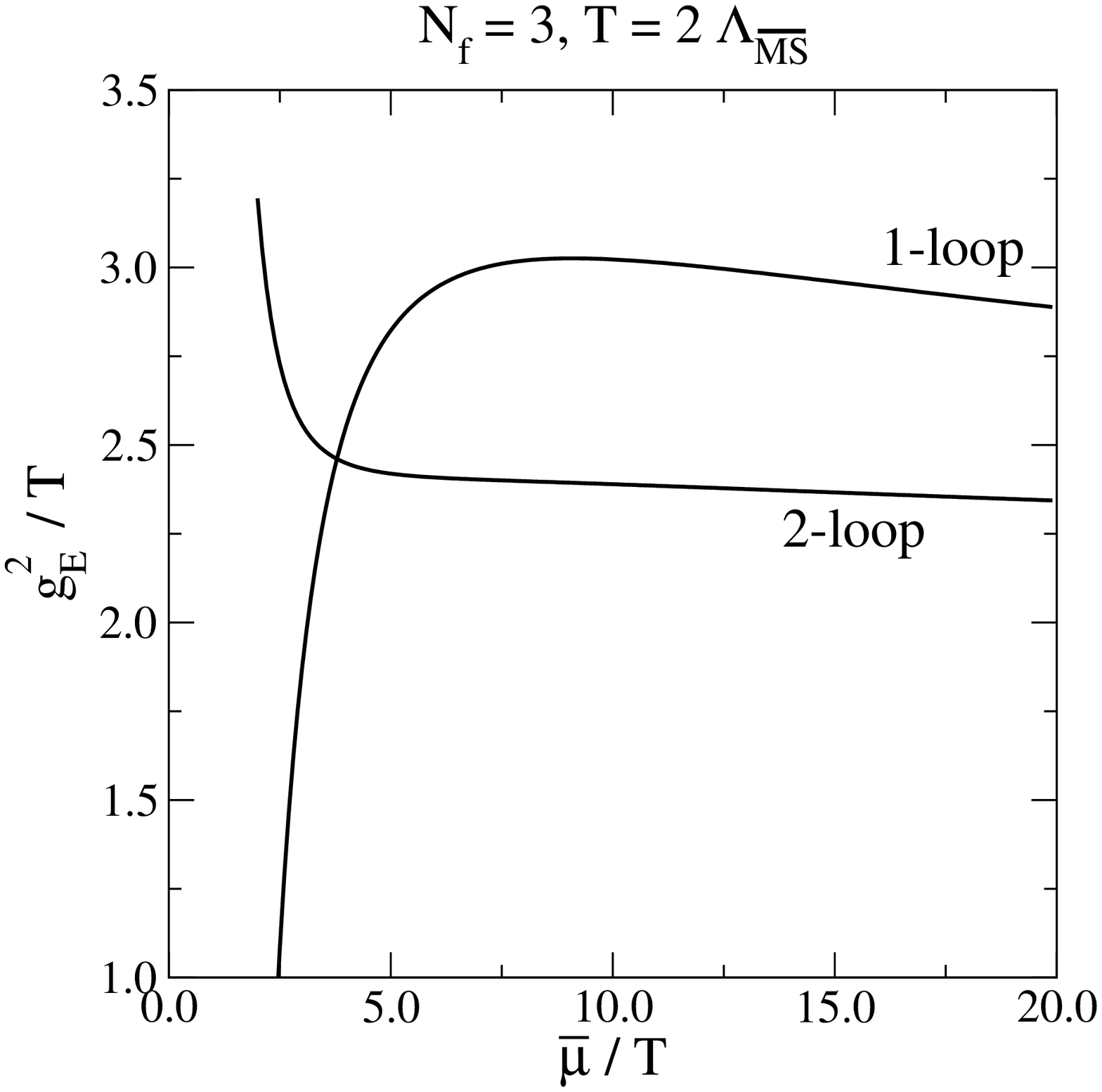}%
}

\caption[a]{\it A comparison of 1-loop 
and 2-loop values for $g_\rmi{E}^2/T$, 
as a function of $\bmu/T$, 
for a fixed $T/\Lambdamsbar =2.0$ and $\Nf = 0,2,3$. } 
\la{fig:mudep}

\end{figure}

Through \eqs\nr{gE1}, \nr{gE2}, \nr{gE3} and \nr{Lambdamsbar}, 
$g_\rmi{E}^2$ is a function $\bmu/T$ and $\bmu/\Lambdamsbar$. The dependence
on $\bmu$ is formally of higher order than the computation. Numerically, 
of course, there is non-vanishing dependence, 
as illustrated in~\fig\ref{fig:mudep}.

\begin{figure}[t]

\centerline{%
\epsfysize=5.0cm\epsfbox{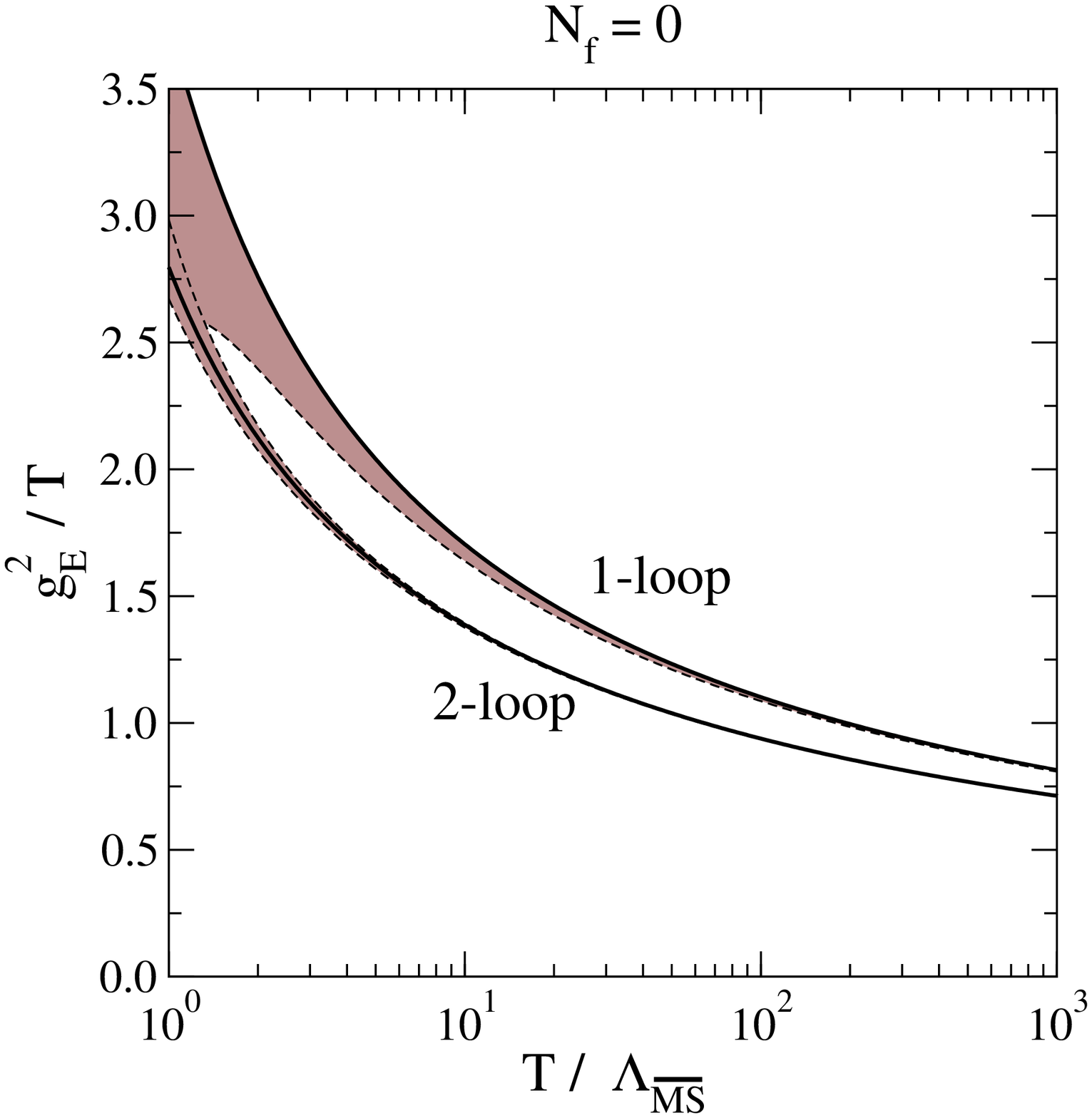}%
~~\epsfysize=5.0cm\epsfbox{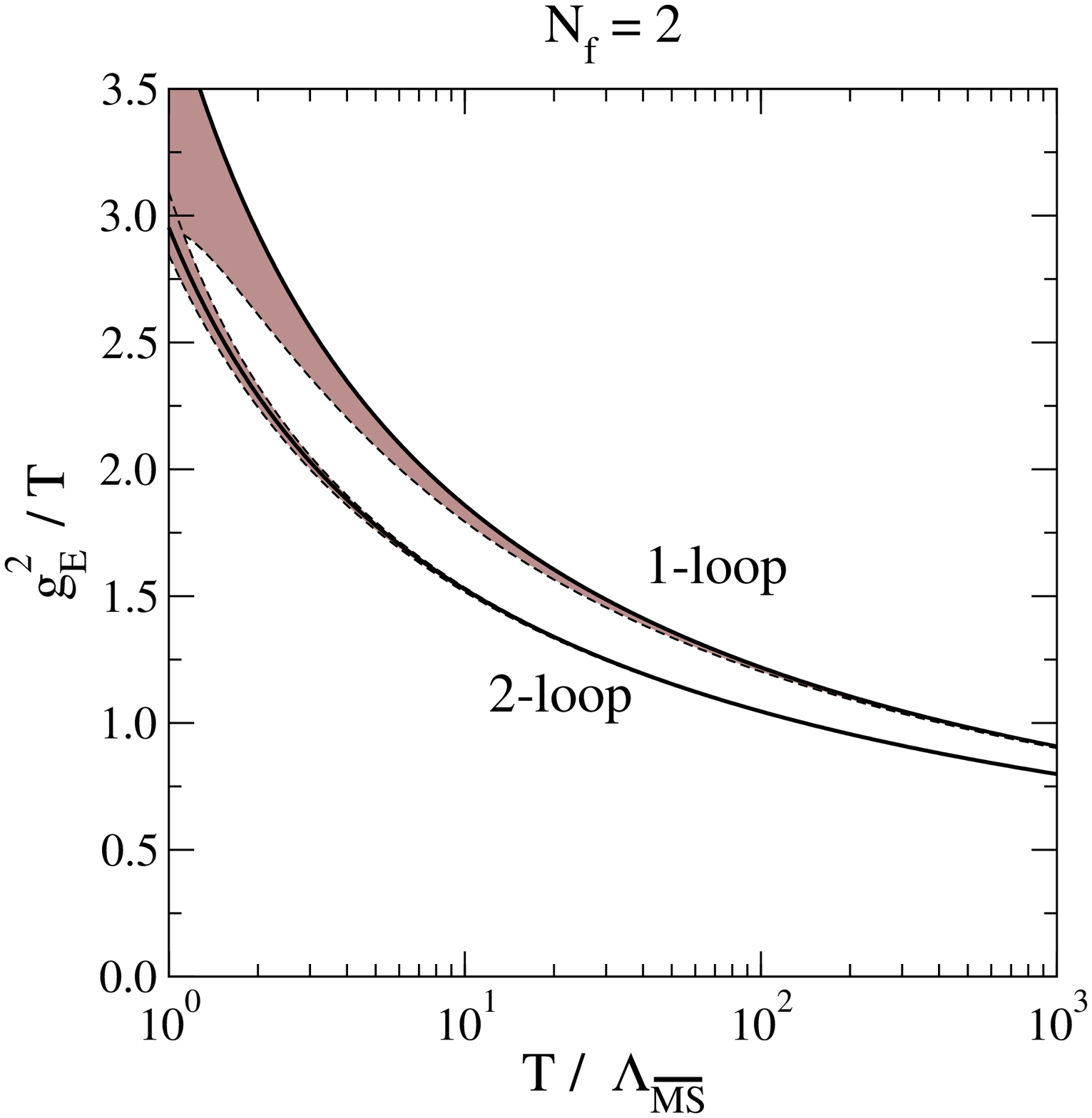}%
~~\epsfysize=5.0cm\epsfbox{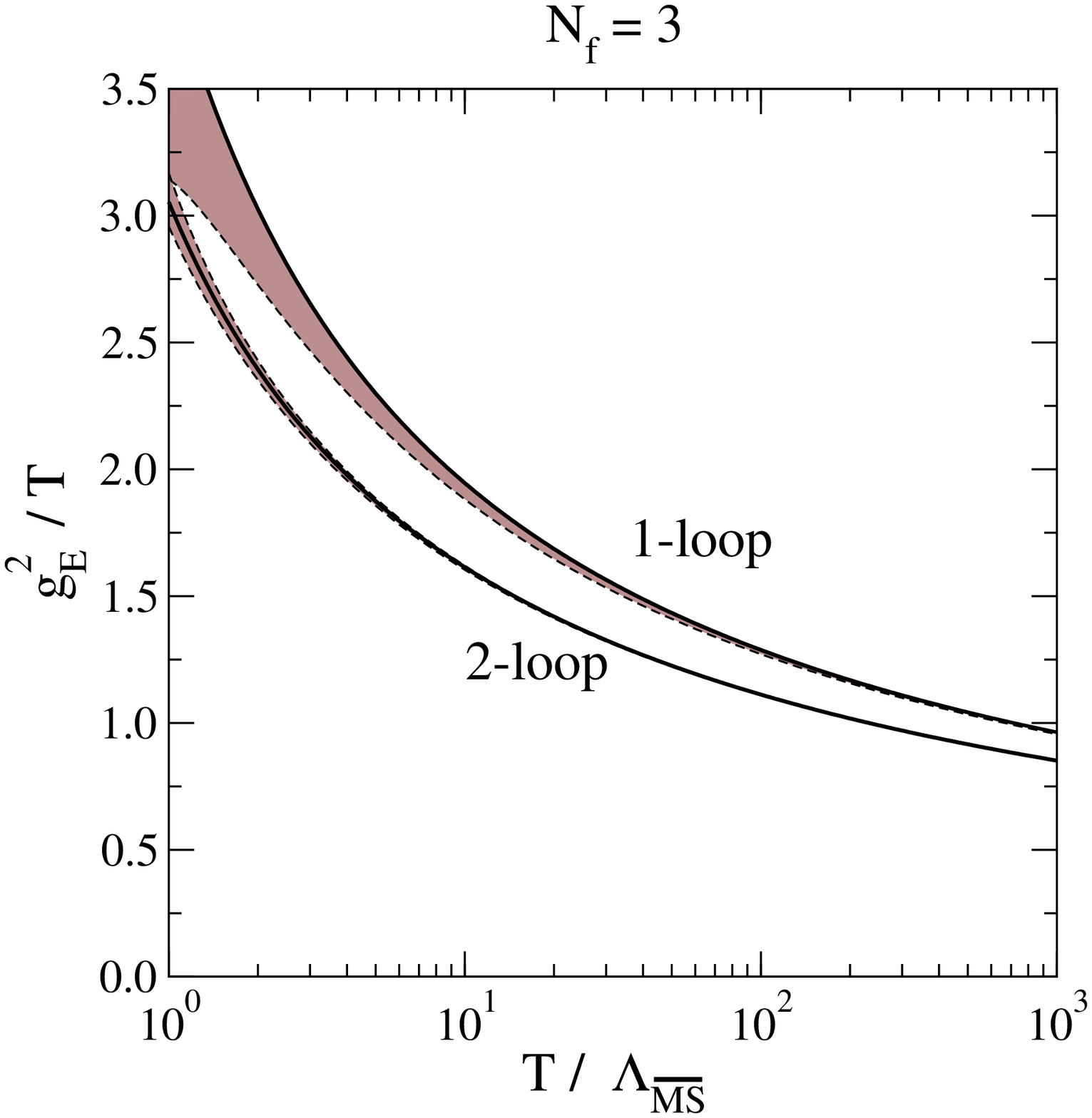}%
}

\caption[a]{\it The 1-loop and 2-loop values for $g_\rmi{E}^2/T$, 
as a function
of $T/\Lambdamsbar$ (solid lines). 
For each $T$ the scale $\bmu$ has been fixed to the 
``principal of minimal sensitivity'' point $\bmu_\rmi{opt}$
following from the 1-loop expression, and varied then in the 
range $\bmu = (0.5 ... 2.0) \times \bmu_\rmi{opt}$ (the grey bands).} 
\la{fig:opt}

\end{figure}

As usual, one may choose some ``optimisation'' criterion which should
lead to a reduced $\bmu$-dependence and thus reasonable convergence. 
We fix $\bmu_\rmi{opt}$ to be the point where 
the 1-loop coupling 
$g_\rmi{E}^2$ has vanishing slope (``principal of minimal
sensitivity''), cf.\ \fig\ref{fig:mudep}, and vary then the scale in the range
$\bmu = (0.5 ... 2.0) \times \bmu_\rmi{opt}$ around this point. Results
are shown in \fig\ref{fig:opt}. The $\bmu$-dependence indeed decreases 
significantly as we go to the 2-loop level. The numerical 2-loop value 
is some 20\% smaller than the 1-loop value. 
It is comforting that the 2-loop value is on the side
to which the ``error band'' of the 1-loop result points, even though it does
not in general lie within that band.

%
\section{Spatial string tension}
\la{se:string}

The computations in the previous sections 
can be given a ``phenomenological''
application, by considering lattice measurements 
of the so-called spatial string
tension.
The spatial string tension is obtained from a rectangular 
Wilson loop $W_s(R_1,R_2)$ in the ($x_1,x_2$)-plane, of 
size $R_1 \times R_2$. The potential $V_s(R_1)$ is defined 
through 
\be
 V_s(R_1) = - \lim_{R_2 \to \infty}
 \frac{1}{R_2} \ln W_s(R_1,R_2) 
 \;, \la{FWilson}
\ee
and the spatial string tension $\sigma_s$ from the asymptotic behaviour 
of the potential, 
\be
 \sigma_s \equiv \lim_{R_1 \to \infty} \frac{V_s(R_1)}{R_1}
 \;. 
\ee 
Since $\sigma_s$ has the dimensionality GeV$^2$, it is often
expressed~\cite{boyd} as the combination
\be
 \frac{\sqrt{\sigma_s}}{T} 
 = \phi \Bigl( \frac{T}{\Tc} \Bigr) 
 \;, \la{sigma_4d}
\ee
where $\phi$ is a (decreasing) dimensionless function, and $\Tc$
is the critical temperature of the deconfinement phase transition.

We now turn to
how the result for $g_\rmi{E}^2$ that we have obtained in this paper, 
combined with other ingredients, allow us to 
obtain an independent prediction for the spatial string tension.

%
\subsection{Three-dimensional prediction}
\la{ss1}

The very same observable as in \eq\nr{FWilson},
exists also in 3d SU(3) gauge theory, 
or ``Magnetostatic QCD''
(MQCD). Since the gauge coupling $g_\rmi{M}^2$ of MQCD is 
dimensionful, $\sigma_s$
must have the form $\sigma_s = c \times g_\rmi{M}^4$, where $c$
is a numerical proportionality constant. It has been determined with 
lattice Monte Carlo methods most recently
in Ref.~\cite{mt} where, after the continuum 
extrapolation, it was expressed as 
\be
 \frac{\sqrt{\sigma_s}}{g_\rmi{M}^2} = 0.553(1) \;.
 \la{sigma_3d}
\ee

In order to compare \eqs\nr{sigma_4d}, \nr{sigma_3d}, we need a relation
between $T$ and $g_\rmi{M}^2$. In the previous section, we obtained 
a relation between $T$ and $g_\rmi{E}^2$. The relation between 
$g_\rmi{E}^2$ and $g_\rmi{M}^2$ is also known, 
up to 2-loop order~\cite{pg}:\footnote{%
 The 2-loop correction 
 $\delta g_\rmii{M}^2 /g_\rmii{E}^2 = 
 - {g_\rmii{E}^2 C_A 
   [2 (C_A C_F + 1) \lambda_\rmii{E}^\rmii{(1)} + 
   (6 C_F - C_A)\lambda_\rmii{E}^\rmii{(2)} ]} / 
   {384 (\pi m_\rmii{E})^2}$
 was ignored in Ref.~\cite{pg}, as it is of higher
 order according to 4d power counting 
 and numerically insignificant.
 }
\be
 g_\rmi{M}^2 = g_\rmi{E}^2 
 \; \biggl[
  1 - \frac{1}{48} \frac{g_\rmi{E}^2 C_A}{\pi m_\rmi{E}}
    - \frac{17}{4608} 
  \biggl( \frac{g_\rmi{E}^2 C_A}{\pi m_\rmi{E}} \biggr)^2
 \biggr]
 \;, \la{gMgE}
\ee
where the 1-loop part was determined already in Ref.~\cite{fkrs}.

It is worth stressing that the corrections 
in \eq\nr{gMgE} are in practice extremely small, even for 
values of $m_\rmi{E}/g_\rmi{E}^2$ corresponding to temperatures
very close to the critical one.
(For $\Nc = 3$ and $\Nf = 0$, 
$(m_\rmi{E}/g_\rmi{E}^2)^2 \approx 0.32\, \log_{10}(T/\Lambdamsbar) + 0.29$.)  
This seems by 
no means obvious {\it a priori}, given the observed slow 
convergence in the case of the vacuum 
energy density of EQCD~\cite{gsixg}. In view of this fact, however, 
we can safely ignore all higher loop corrections in~\eq\nr{gMgE}.

Another source of errors in going from EQCD to MQCD are the
higher order operators that have been truncated from the action of MQCD. 
As discussed in Ref.~\cite{gsixg}, they are expected to contribute at the 
relative order $\mathcal{O}(g_\rmi{E}^6/m_\rmi{E}^3)$, i.e.\ at the same
order that 3-loop corrections enter~\eq\nr{gMgE}. From this 
consideration, one might expect them to again be numerically
negligible. In principle one could avoid this assumption, however: 
the ratio $\sqrt{\sigma_s}/g_\rmi{E}^2$ has been estimated 
in Ref.~\cite{mu} through direct numerical simulations in EQCD. 
Unfortunately  the statistical and 
particularly the systematic errors appear to be non-vanishing (no continuum 
extrapolation was carried out for this quantity), so that 
we prefer to follow the line starting from 
\eq\nr{sigma_3d} in the following. Nevertheless it would be interesting
to learn more about the importance of the higher order operators.

Now, as we know $g_\rmi{E}^2/T$ as a function of $T/\Lambdamsbar$
from \fig\ref{fig:opt}, \eqs\nr{sigma_3d} and 
\nr{gMgE} allow us to obtain $\sqrt{\sigma_s}/T$ as a function of 
the same variable.  In order to compare 
with \eq\nr{sigma_4d}, however, 
we still need to relate $\Lambdamsbar$ to $\Tc$. 
This problem has also been addressed with 4d lattice simulations, 
as we review in \se\ref{se:Tc}.

%
\subsection{Critical temperature in ``perturbative units''}
\la{se:Tc}

The determination of $\Tc / \Lambdamsbar$ is a classic 
problem in lattice QCD. Two main lines have been 
followed, one
going via the zero temperature string tension $\sqrt{\sigma}$, 
the other via the Sommer scale $r_0$~\cite{rs}.

Values obtained for $\Tc/\sqrt{\sigma}$ by various
lattice collaborations are summarised 
in Ref.~\cite{hn}, Table 7. Traditionally the values were around
$\Tc/\sqrt{\sigma} = 0.630(5)$~\cite{bb}, but Ref.~\cite{hn} argues
in favour of a slightly larger number in the continuum limit. Indeed 
the most precise estimate appears to come from 
Ref.~\cite{ltw}, where $\Tc/\sqrt{\sigma} = 0.646(3)$ is cited.
Combining with $\Lambdamsbar / \sqrt{\sigma} = 0.555(19)$ from
Ref.~\cite{bs}, we are lead to 
\be
 \frac{\Tc}{\Lambdamsbar} = 1.16(4) \;. \la{Tc_sigma}
\ee
The error is dominated by the one in $\Lambdamsbar / \sqrt{\sigma}$.

A value for $r_0 \Tc$, on the other hand, has been obtained 
in Ref.~\cite{sn}: $r_0 \Tc = 0.7498(50)$. Combining with 
$r_0 \Lambdamsbar = 0.602(48)$ from Ref.~\cite{al}
(the value $r_0 \Lambdamsbar = 0.586(48)$ from a few lines below Eq.~(4.11)
in Ref.~\cite{ns} is well within error bars), one obtains
\be
 \frac{\Tc}{\Lambdamsbar} = 1.25(10) \;. \la{Tc_r0}
\ee
This is consistent, within statistical errors, with \eq\nr{Tc_sigma}, 
if favouring a slightly larger central value. Again the error is 
dominated by the zero-temperature part, $r_0 \Lambdamsbar$ in 
this case. In general it might be expected, though, that 
systematic uncertainties are better under control 
in the extraction of $r_0$ than of $\sqrt{\sigma}$, since the static
potential needs to be computed only up to intermediate distances. 

Apart from going through $\sqrt{\sigma}$ and $r_0$, there is also 
a third possibility~\cite{gupta}. It is based on directly determining 
a (lattice) $\Lambda$-parameter from the scaling of a suitably defined 
renormalised gauge coupling at the critical point, and converting at the
end to the $\msbar$ scheme. The value obtained is
\be
 \frac{\Tc}{\Lambdamsbar} = 1.15(5) \;, \la{Tc_gupta}
\ee
consistent with \eqs\nr{Tc_sigma} and \nr{Tc_r0}.

To be conservative, we will consider the interval 
$\Tc/\Lambdamsbar = 1.10...1.35$ in the following, 
encompassing the central values as well as the error 
bars of \eqs\nr{Tc_sigma}--\nr{Tc_gupta}.

%
\subsection{Four-dimensional measurement}

The spatial string tension of 4d pure SU(3) gauge theory 
at temperatures above the critical one, as a function of $T/\Tc$,
has been measured at $N_\tau = 8$ in Ref.~\cite{boyd} (cf.\ Fig.~11).
There are, of course, systematic uncertainties, both from the lack 
of a continuum
extrapolation as well as from how the string tension is extracted by 
fitting to the large-distance behaviour of the static potential. 
Nevertheless, we expect that
the results are in the right ballpark. 

\begin{figure}[t]

\centerline{%
\epsfysize=8.0cm\epsfbox{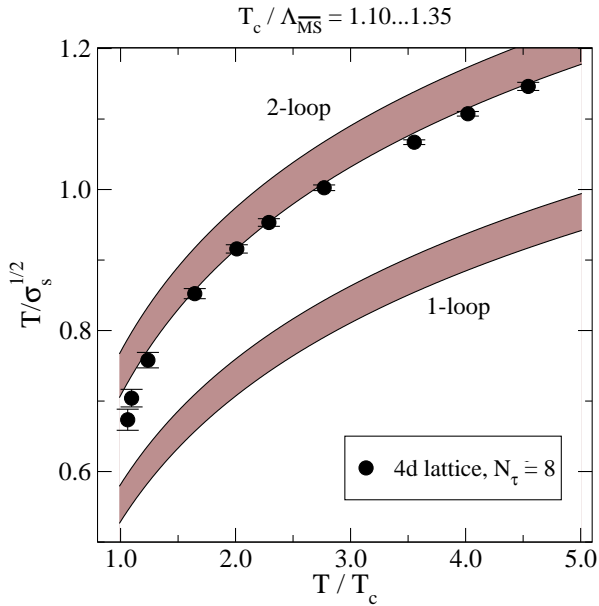}%
}

\caption[a]{\it We compare 4d lattice data for the spatial string tension, 
taken from Ref.~\cite{boyd}, with expressions obtained by combining 1-loop
and 2-loop results for $g_\rmi{E}^2$ together with \eq\nr{gMgE} and
the non-perturbative
value of the string tension of 3d SU(3) gauge theory, \eq\nr{sigma_3d}.
The upper edges of the bands correspond to $\Tc/\Lambdamsbar = 1.35$,
the lower edges to $\Tc/\Lambdamsbar = 1.10$.} 

\la{fig:compare}
\end{figure}

Given the considerations in Secs.~\ref{ss1}, \ref{se:Tc}, we can thus 
compare the 3d and the 4d determinations of $\sqrt{\sigma_s}/T$.
The result is shown in \fig\ref{fig:compare}, 
where $T/\sqrt{\sigma_s}$ is plotted.
We observe a significant discrepancy at 1-loop level (as most 
recently pointed out in Ref.~\cite{pg2}), but a remarkable
agreement once we go to 2-loop level. It is also noteworthy 
that the functional form of the 2-loop curve appears to 
match the behaviour of the lattice data down to low temperatures.

%
\section{Conclusions}
\la{se:concl}

The main purpose of this paper has been the analytic computation
of the 2-loop effective gauge coupling of QCD at finite temperatures, 
defined as a matching coefficient appearing in the dimensionally reduced
effective theory, EQCD.\footnote{%
 Other ``effective gauge couplings'' can of course also be defined; 
 for a recent review, see Ref.~\cite{kz}. The difference is that in 
 these cases all momentum scales influence the effective gauge coupling, 
 so that perturbation theory cannot be reliably applied
 for its computation. 
 } 
The result is given in \eqs\nr{gE1}--\nr{gE3}.
We have also determined a new contribution of order $\mathcal{O}(g^6T^4)$
to the pressure of hot QCD; the information is contained
in \eq\nr{bE3}, and how it 
enters the pressure is explained in Ref.~\cite{gsixg}.

The 2-loop correction we find is numerically substantial, 
some 20\% of the 1-loop expression. 
This indicates that while perturbation theory 
is in principle still under control, 
if restricted to the parametrically hard modes $p\sim 2\pi T$ only, 
it is important to push it to
a sufficiently high order, in order to obtain precise results. 

Our expression for the effective gauge coupling 
has a direct ``phenomenological'' application, 
in that it allows for a parameter-free comparison of 3d MQCD 
and 4d full theory 
results for an observable called the spatial string tension. We find
that the 2-loop correction computed here improves the match
between the two results quite significantly, 
down to temperatures very close to the critical one.
A small discrepancy still remains but, given
that no continuum extrapolation was taken in 4d lattice simulations, 
that the extraction of the spatial
string tension may involve systematic uncertainties 
due to large subleading terms in the $r$-dependence
of the spatial static potential $V_s(r)$~\cite{lw2}, and that  
there also has to be some room for residual 3-loop corrections, as well as
improvements in the matching between EQCD and MQCD, 
we do not consider this discrepancy to be worrying. We do believe 
that the discrepancy can be decreased by improving systematically
on the various ingredients that enter the comparison. 

These conclusions support a picture of thermal QCD according to which
the {\em parametrically} ``hard'' scales, $p\sim 2\pi T$, can be treated 
perturbatively, almost as soon as we are in the deconfined phase, will the 
{\em parametrically} ``soft'' scales, $p\sim gT,g^2T$, require in general 
a non-perturbative analysis within one of the effective theories 
describing their dynamics. For the observable we considered here, 
in fact, even the colour-electric scale $p \sim gT$ could be integrated
out perturbatively, but it is known that this is 
in general not the case. We should like to stress that this conclusion
is rather non-trivial, as there {\em numerically} is little hierarchy
between the scales $2\pi T, gT, g^2 T$ 
at the realistic temperatures that we have been considering. 

%
\section*{Acknowledgements}

We acknowledge useful discussions with P.~Giovannangeli 
and C.P.~Korthals Altes, and thank E.~Laermann for providing the 
lattice data for the spatial string tension from Ref.~\cite{boyd}.



\appendix
\renewcommand{\thesection}{Appendix~\Alph{section}}
\renewcommand{\thesubsection}{\Alph{section}.\arabic{subsection}}
\renewcommand{\theequation}{\Alph{section}.\arabic{equation}}


%
\section{Expansions for master integrals}

Using the notation introduced in the text, the master integrals
of \eq\nr{Imaster} read, up to $\mathcal{O}(\epsilon)$:
\ba
 I_\rmi{b}(1) \!\! & = & \!\!
 \mu^{-2\epsilon} \frac{T^2}{12}
 \biggl\{
  1 + \epsilon \biggl[ 
   2 \ln \biggl( \frac{\bmu e^{\gamma_\rmi{E}}}{4\pi T} \biggr) 
 + 2 - 2 \gamma_\rmi{E} + 
   2 \frac{\zeta'(-1)}{\zeta(-1)}
 \biggr] 
 \biggr\} 
 \;, \\ 
 I_\rmi{b}(2) \!\! & = & \!\! 
 \mu^{-2 \epsilon} \frac{1}{(4\pi)^2}
 \biggl\{
  \frac{1}{\epsilon} + 2 \ln \biggl(\frac{\bmu e^{\gamma_\rmi{E}}}{4\pi T}
 \biggr) 
 + \epsilon \biggl[ 
   2 \ln^2 \biggl( \frac{\bmu e^{\gamma_\rmi{E}}}{4\pi T} \biggr)
   + \frac{\pi^2}{4} - 2 \gamma_\rmi{E}^2 - 4 \gamma_1 
 \biggr] 
 \biggr\}
 \;, \\
 I_\rmi{b}(3) \!\! & = & \!\!
 \mu^{-2\epsilon} \frac{\zeta(3)}{128 \pi^4 T^2}
 \biggl\{
  1 + \epsilon \biggl[ 
   2 \ln \biggl( \frac{\bmu e^{\gamma_\rmi{E}}}{4\pi T} \biggr) 
 + 2 - 2 \gamma_\rmi{E} + 
   2 \frac{\zeta'(3)}{\zeta(3)}
 \biggr] 
 \biggr\} 
 \;, \\ 
 I_\rmi{f}(1) \!\! & = & \!\!
 \mu^{-2\epsilon} \biggl( - \frac{T^2}{24} \biggr)
 \biggl\{
  1 + \epsilon \biggl[ 
   2 \ln \biggl( \frac{\bmu e^{\gamma_\rmi{E}}}{\pi T} \biggr) 
 + 2 - 6 \ln 2 - 2 \gamma_\rmi{E} 
    + 
   2 \frac{\zeta'(-1)}{\zeta(-1)}
 \biggr] 
 \biggr\} 
 \;, \\ 
 I_\rmi{f}(2) \!\! & = & \!\! 
 \mu^{-2 \epsilon} \frac{1}{(4\pi)^2}
 \biggl\{
  \frac{1}{\epsilon} + 2 \ln \biggl(\frac{\bmu e^{\gamma_\rmi{E}}}{\pi T}
 \biggr) 
 + \epsilon \biggl[ 
   2 \ln^2 \biggl( \frac{\bmu e^{\gamma_\rmi{E}}}{\pi T} \biggr)
   + \frac{\pi^2}{4}  -4 \ln^2 2 - 2 \gamma_\rmi{E}^2 - 4 \gamma_1 
 \biggr] 
 \biggr\}, \hspace*{1.0cm} \\
 I_\rmi{f}(3) \!\! & = & \!\!
 \mu^{-2\epsilon} \frac{7 \zeta(3)}{128 \pi^4 T^2}
 \biggl\{
  1 + \epsilon \biggl[ 
   2 \ln \biggl( \frac{\bmu e^{\gamma_\rmi{E}}}{\pi T} \biggr) 
 + 2 
   - \frac{12}{7} \ln 2 - 2 \gamma_\rmi{E} + 
   2 \frac{\zeta'(3)}{\zeta(3)}
 \biggr] 
 \biggr\} 
 \;.
\ea



\begin{thebibliography}{99}

\bibitem{uh}
U.W.~Heinz,
AIP Conf.\ Proc.\  {739} (2005) 163
[nucl-th/0407067].

\bibitem{es}
E.V.~Shuryak,
Sov.\ Phys.\ JETP {47} (1978) 212
[Zh.\ Eksp.\ Teor.\ Fiz.\  {74} (1978) 408];
S.A.~Chin,
Phys.\ Lett.\ B {78} (1978) 552.

\bibitem{az}
P.~Arnold and C.~Zhai,
Phys.\ Rev.\  {D 50} (1994) 7603
[hep-ph/9408276];
%
{\it ibid.}\  {51} (1995) 1906
[hep-ph/9410360].

\bibitem{zk}
C.~Zhai and B.~Kastening,
Phys.\ Rev.\  {D 52} (1995) 7232 [hep-ph/9507380].

\bibitem{jk}
J.I.~Kapusta,
Nucl.\ Phys.\ B {148} (1979) 461.

\bibitem{linde}
A.D.~Linde,
Phys.\ Lett.\ {B 96} (1980) 289.

\bibitem{gpy}
D.J.~Gross, R.D.~Pisarski and L.G.~Yaffe,
Rev.\ Mod.\ Phys.\ {53} (1981) 43.

\bibitem{bn}
E. Braaten and A. Nieto,
Phys.\ Rev.\ D 53 (1996) 3421 [hep-ph/9510408].

\bibitem{adjoint}
K.~Kajantie, M.~Laine, K.~Rummukainen and M.~Shaposhnikov,
Nucl.\ Phys.\ B {503} (1997) 357
[hep-ph/9704416].

\bibitem{gsixg}
K.~Kajantie, M.~Laine, K.~Rummukainen and Y.~Schr\"oder,
Phys.\ Rev.\ D 67 (2003) 105008
[hep-ph/0211321]. 

\bibitem{dr}
P. Ginsparg, 
Nucl.\ Phys.\ B 170 (1980) 388;
%
T. Appelquist and R.D. Pisarski,
Phys.\ Rev.\ D 23 (1981) 2305.

\bibitem{generic}
K.~Kajantie, M.~Laine, K.~Rummukainen and M.~Shaposhnikov,
Nucl.\ Phys.\ {B 458} (1996) 90 [hep-ph/9508379].

\bibitem{bp}
R.D.~Pisarski,
Phys.\ Rev.\ Lett.\  {63} (1989) 1129;
%
E.~Braaten and R.D.~Pisarski,
Phys.\ Rev.\ D {45} (1992) 1827.

\bibitem{own}
M.~Laine,
hep-ph/0301011.

\bibitem{owe}
E.~Laermann and O.~Philipsen,
Ann.\ Rev.\ Nucl.\ Part.\ Sci.\  {53} (2003)  163
[hep-ph/0303042].

\bibitem{chris}
C.P.~Korthals Altes,
hep-ph/0308229.

\bibitem{mu}
M.~Laine and O.~Philipsen,
Phys.\ Lett.\ B {459} (1999) 259
[hep-lat/9905004];
%
A.~Hart and O.~Philipsen,
Nucl.\ Phys.\ B {572} (2000) 243
[hep-lat/9908041];
%
A.~Hart, M.~Laine and O.~Philipsen,
Nucl.\ Phys.\ B {586} (2000) 443
[hep-ph/0004060].


\bibitem{plaq}
A.~Hietanen, K.~Kajantie, M.~Laine, K.~Rummukainen and Y.~Schr\"oder,
JHEP 01 (2005) 013 
[hep-lat/0412008].

\bibitem{mv}
M.~Laine and M.~Veps\"al\"ainen,
JHEP {02} (2004) 004 
[hep-ph/0311268].

\bibitem{ag}
P.~Giovannangeli and C.P.~Korthals Altes,
hep-ph/0212298;
%
hep-ph/0412322.

\bibitem{bmp}
P.~Bialas, A.~Morel, B.~Petersson, K.~Petrov and T.~Reisz,
Nucl.\ Phys.\ B {581} (2000) 477
[hep-lat/0003004];
%
Nucl.\ Phys.\ B {603} (2001) 369
[hep-lat/0012019].

\bibitem{boyd}
G.~Boyd, J.~Engels, F.~Karsch, E.~Laermann, C.~Legeland, 
M.~L\"utgemeier and B.~Petersson,
Nucl.\ Phys.\ B {469} (1996) 419
[hep-lat/9602007].

\bibitem{pg2}
P.~Giovannangeli,
hep-ph/0410346.

\bibitem{mt}
M.J.~Teper,
Phys.\ Rev.\ D {59} (1999) 014512
[hep-lat/9804008];
%
B.~Lucini and M.~Teper,
Phys.\ Rev.\ D {66} (2002) 097502
[hep-lat/0206027].

\bibitem{pg}
P.~Giovannangeli,
Phys.\ Lett.\ B {585} (2004)  144
[hep-ph/0312307].

\bibitem{fkrs}
K.~Farakos, K.~Kajantie, K.~Rummukainen and M.E.~Shaposhnikov,
Nucl.\ Phys.\ B {425} (1994)  67
[hep-ph/9404201].

\bibitem{sc}
S.~Chapman,
Phys.\ Rev.\ D {50} (1994) 5308
[hep-ph/9407313].

\bibitem{mrs}
E.~Meg\'{\i}as, E.~Ruiz Arriola and L.L.~Salcedo,
Phys.\ Rev.\ D {69} (2004)  116003
[hep-ph/0312133].

\bibitem{do}
D.~Diakonov and M.~Oswald,
Phys.\ Rev.\ D {70} (2004)  105016
[hep-ph/0403108].

\bibitem{parity}
K.~Kajantie, M.~Laine, K.~Rummukainen and M.E.~Shaposhnikov,
Phys.\ Lett.\ B {423} (1998)  137
[hep-ph/9710538].

\bibitem{lfa}
L.F.~Abbott,
Nucl.\ Phys.\ B {185} (1981) 189.

\bibitem{lw}
M.~L\"uscher and P.~Weisz,
Nucl.\ Phys.\ B {452} (1995) 213
[hep-lat/9504006].

\bibitem{laporta}
S.~Laporta,
Int.\ J.\ Mod.\ Phys.\ A {15} (2000) 5087
[hep-ph/0102033].

\bibitem{ysproc}
Y.~Schr\"oder, 
Nucl.\ Phys.\ B (Proc.\ Suppl.)\ 116 (2003) 402
[hep-ph/0211288].

\bibitem{ae}
P.~Arnold and O.~Espinosa,
Phys.\ Rev.\ D {47} (1993) 3546
[hep-ph/9212235]; 
{\it ibid.}\ D {50} (1994) 6662 (Erratum).

\bibitem{hl}
S.~Huang and M.~Lissia,
Nucl.\ Phys.\ B {438} (1995) 54
[hep-ph/9411293].

\bibitem{ysproc2}
Y.~Schr\"oder, 
hep-ph/0410130.

\bibitem{rs}
R.~Sommer,
Nucl.\ Phys.\ B {411} (1994)  839
[hep-lat/9310022].

\bibitem{hn}
M.~Hasenbusch and S.~Necco,
JHEP {08} (2004) 005
[hep-lat/0405012].

\bibitem{bb}
B.~Beinlich, F.~Karsch, E.~Laermann and A.~Peikert,
Eur.\ Phys.\ J.\ C {6} (1999) 133
[hep-lat/9707023].

\bibitem{ltw}
B.~Lucini, M.~Teper and U.~Wenger,
JHEP {01} (2004)  061
[hep-lat/0307017].

\bibitem{bs}
G.S.~Bali and K.~Schilling,
Phys.\ Rev.\ D {47} (1993)  661
[hep-lat/9208028].

\bibitem{sn}
S.~Necco,
Nucl.\ Phys.\ B {683} (2004) 137
[hep-lat/0309017].

\bibitem{al}
S.~Capitani, M.~L\"uscher, R.~Sommer and H.~Wittig  [ALPHA Collaboration],
Nucl.\ Phys.\ B {544} (1999)  669
[hep-lat/9810063].

\bibitem{ns}
S.~Necco and R.~Sommer,
Nucl.\ Phys.\ B {622} (2002) 328
[hep-lat/0108008].

\bibitem{gupta}
S.~Gupta,
Phys.\ Rev.\ D {64} (2001) 034507
[hep-lat/0010011].

\bibitem{kz}
O.~Kaczmarek and F.~Zantow,
hep-lat/0502012.

\bibitem{lw2}
M.~L\"uscher and P.~Weisz,
JHEP {07} (2002) 049
[hep-lat/0207003].


\end{thebibliography}
\end{document}